\documentclass{PoS}

\usepackage{verbatim}

\newcommand{\ApJ}{ApJ}
\newcommand{\AeA}{A\&A}
\newcommand{\PRL}{PRL}
\newcommand{\PRD}{PRD}
\newcommand{\etal}{et alii}

\newcommand{\AMS}{\textsf{AMS}}
 
\newcommand{\ie}{\textit{i.e.}}

\def\Journal#1#2#3#4{{#4}, {#1}, {#2}, #3} 
\newcommand{\citep}{\cite}

\newcommand{\p}{\textrm{\ensuremath{p}}}
\newcommand{\Hyd}{\textrm{H}}
\newcommand{\He}{\textrm{He}}
\newcommand{\Li}{\textrm{Li}}
\newcommand{\Be}{\textrm{Be}}
\newcommand{\B}{\textrm{B}}
\newcommand{\C}{\textrm{C}}
\newcommand{\N}{\textrm{N}}
\newcommand{\Oxy}{\textrm{O}}
\newcommand{\Si}{\textrm{Si}}
\newcommand{\Fe}{\textrm{Fe}}

\newcommand{\BC}{\textrm{B}/\textrm{C}}

\newcommand{\Htwo}{\ensuremath{^{2}}\textrm{H}}
\newcommand{\Het}{\ensuremath{^{3}}\textrm{He}}

\newcommand{\pbarp}{\textrm{\ensuremath{\bar{p}/p}}}
\newcommand{\pbar}{\textrm{\ensuremath{\bar{p}}}}
\newcommand{\nbar}{\textrm{\ensuremath{\bar{n}}}}
\newcommand{\dbar}{\ensuremath{\rm \overline{d}}}
\newcommand{\tbar}{\ensuremath{\rm \overline{t}}}
\newcommand{\hebar}{\ensuremath{\rm \overline{He}}}
\newcommand{\hetbar}{\ensuremath{\rm \overline{^{3}He}}} 
\newcommand{\htbar}{\ensuremath{\rm \overline{^{3}H}}}

\usepackage{xspace}

\title{Secondary antinuclei from supernova remnants and background for dark matter searches}
\ShortTitle{Acceleration and production of CR Antinuclei in SNRs}

\author{\speaker{Nicola Tomassetti}\thanks{E-mail: {nicola.tomassetti@cern.ch}}\\
Department of Physics and Earths Science, Universit{\`a} di Perugia, and INFN-Perugia, I-06100 Perugia, Italy\\
}
\author{{Alberto Oliva}\thanks{E-mail: {alberto.oliva@cern.ch}}\\
Centro de Investigaciones Energ{\'e}ticas, Medioambientales y Tecnol{\'o}gicas CIEMAT, E-28040 Madrid, Spain\\
}

\abstract{
  We compute the spectra of cosmic-ray (CR) nuclei and anti-nuclei under a scenario where 
  hadronic interaction processes inside supernova remnants (SNRs) can produce a diffusively-shock-accelerated 
  ``source component'' of secondary particles. This scenario is able to explain the recent measurements reported 
  by AMS on the antiproton/proton ratio, that is found to be remarkably constant at $\sim$\,60-450GeV of kinetic energy. 
  However, as we will show, this explanation is ruled out by the new AMS data on the B/C ratio, which is found to decrease 
  steadily up to TeV/n energies. With the constraints provided by the two ratios, we calculate conservative (B/C driven) 
  and speculative ($\bar{p}/p$ driven) SNR-induced flux contribution for the spectra of antideuteron and antihelium in CRs, 
  along with their standard secondary component expected from CR collisions in the interstellar gas. 
  We found that the SNR component of anti-nuclei can be significantly large at high-energy, above a few $\sim$\,10 GeV/n, 
  but it is always sub-dominant at sub-GeV/n energies, that is, the energy region where dark-matter induced signals may 
  exceed the standard astrophysical background. Furthermore, the total antinuclei flux from insterstellar spallation 
  plus SNR-component is tightly bounded by the data, so that hadronic production in SNRs has a minor impact on 
  the astrophysical background for dark matter searches.
}

\FullConference{35th International Cosmic Ray Conference -- ICRC2017 --\\
		10-20 July, 2017\\
		Bexco, Busan, Korea}

\begin{document}

\section{Introduction}

Light antinuclei in cosmic rays (CRs) such as antiproton (\pbar), antideuteron (\dbar), and antihelium (\hebar)
are excellent messengers for the search of dark-matter (DM) in the Galaxy.
For a wide range of DM mass, the annihilation of decay of DM particles may generate antinucleons, \pbar{} or \nbar,
which can merge into antinuclei (\dbar, \htbar, \hetbar) thereby providing signatures in the energy window below a few GeV \citep{Aramaki2016,Cirelli2014,Fornengo2017}. 
The antiproton-to-proton (\pbarp) ratio in CRs has been measured by the Alpha Magnetic Spectrometer (\AMS)
experiment from $\sim$\,0.5 to 450\,GeV of kineti cenergy \citep{Aguilar2016PbarP}.
At $E\gtrsim$\,60\,GeV, the \pbarp{} ratio is found to be unexpectedly constant with energy,
in contrast to  standard-model calculations of secondary \pbar{} production in the interstellar matter (ISM),
from which the \pbarp{} ratio is expected to decrease with energy.
Due to this tension, several authors claimed the need of assessing the astrophysical antimatter background \citep{Salati2016,Lipari2017,Kohri2016,Feng2016,Giesen2015}. 
While interpretations in terms of DM have been proposed \cite{Chen2015}, it was also suggested that a high-energy antiproton ``excess''
may arise from hadronic interaction processes inside supernova remnants (SNRs) \citep{BlasiSerpico2009,MertschSarkar2014}. 
A constant \pbarp{} ratio can in fact arise from a source antiproton component.
Calculations in this direction have been performed recently \citep{TomassettiFeng2017,Herms2017,Cholis2017}.
In contrast, \dbar{} or \hebar{} nuclei have never been observed in CRs,
yet detection experiments have established tight upper limits to the flux of these particles \citep{Aramaki2016}.
The search for antinuclei in CRs is progressing rapidly: \AMS{} is gradually approaching the expected level 
of astrophysical background \citep{Kounine2011}, while the first science flight of the GAPS project will be held in the coming years  \citep{GAPS}.
Recently, hints for \hebar{} events may have been identified by \AMS, strengthening the idea of a possible antimatter excess in CRs \citep{Sokol2017}.
Hence the first observation of antinuclei in CRs has concrete chances to be achieved by the incoming generation of CR detection experiments.

In this paper, we present flux calculations of CR nuclei and antinuclei under a model of acceleration
and propagation which accounts for the production of secondary particles inside SNRs.
Our calculations are carried out within the linear diffusive-shock-acceleration (DSA) theory and the two-halo model of diffusive propagation.
Our statistical analysis are based on standard  $\chi^{2}$-techniques using new experimental data on 
the boron-to-carbon (\BC) and \pbarp{} ratios released by \AMS{} in the GeV-TeV energy region.
As we will show, the two ratios support well the occurrence of secondary production processes inside SNRs.
However, \BC-driven and \pbar-driven constraints give inconsistent results 
in terms of model parameters describing production and acceleration of CR particles.
Using both ratios separately in order to calibrate the model, we derive
conservative (\ie, \BC{}-driven) and speculative (\pbarp{}-driven) 
predictions for the \dbar{} and \hebar{} fluxes in CRs.
We discuss our results in the context of dark matter searches in space.

\section{Calculations}

Secondary production calculations involve a large number of CR+ISM reactions, 
describing the generation of secondary particles and antiparticles from the fragmentation of CR nuclei off
the gas (interstellar or circumstellar) composed by hydrogen and helium \citep{Grenier2015}. 
We account for the production of several isotopes such as  \Htwo, \Het, $^{6,7}$\Li, $^{7,9,10}$\Be, and $^{10,11}$\B{} from the disintegration
of \C-\N-\Oxy, \Si, and \Fe, using cross-section formulae
evaluated in \cite{Tomassetti2012Iso,TomassettiFeng2017} (for the production of \Hyd-\He-isotopes)
and in \cite{Tomassetti2015XS} (for \Li-\Be-\B{} isotopes). 
We therefore account for production of antinuclei such as \pbar, \nbar, \dbar, \tbar{} and  \hebar{} from fragmentation of CR proton and helium.
The production of antinucleons is modeled using the semi-analytical parameterization of \cite{DiMauro2014}, which gives the Lorentz-invariant
distribution function, $f^{CR\rightarrow\,\bar{p,n}}_{\rm ISM} \equiv E_{\bar{p}}\frac{d^{3}\sigma}{dp^{3}_{\bar{p,n}}}$, for all relevant \p-\p{} collisions.
Dedicated parameterizations \cite{InelasticXS} were used to handle other CR+ISM collisions (\p-\He, \He-\p, \He-\He).
The production of antinuclei such as \dbar=\{\nbar,\pbar\} and \hebar=\{\nbar,\pbar,\pbar\} has been re-evaluated within
an improved formulation of the nuclear coalescence model,
based on \citep{Donato2008}.
According to this model,
when \pbar{} or \nbar{} are produced in hadronic jets of nucleus-nucleus collisions,
an antinucleus is formed when the relative momenta of all pairs of antinucleons
lie within the coalescence momentum $p_{0}$.
Using accelerator data \citep{DbarXS}, we set $p_{0}\cong$\,90\,MeV/c, 
The production of \tbar=\{\nbar,\nbar,\pbar\} is also included in our calculations of the total \hebar{} flux.
Destruction reaction and tertiary productions are accounted as well.
The CR acceleration spectrum is calculated using the linear diffusive-shock-acceleration (DSA) theory,
where we account for production and destruction processes of secondary and tertiary fragments. 
Similar calculations are done in earlier works for CR nuclei \citep{MertschSarkar2014,CholisHooper2014,TomassettiDonato2012} 
and antiprotons \citep{BlasiSerpico2009,Herms2017}.
Here we have followed the formalism in \citep{TomassettiDonato2012} as generalized in \citep{TomassettiFeng2017}
and now extended to DSA acceleration of antinuclei.
The amount of secondary production during acceleration is regulated by the product $\tau_{\rm snr}n_{-}$ between SNR age and SNR gas density; 
The subsequent propagation of CRs in the ISM is described under a two-halo model of CR diffusion and nuclear interactions \citep{Tomassetti2015TwoHalo,GuoJin2016}.
The size of the two halos is defined by $l=$\,0.5\,kpc and $L=$\,5\,kpc.
The rigidity ($R$) dependent diffusion coefficient in proximity of the disk ($|z|<l$) is $K\propto \beta K_{0}(R/{\rm GV})^{\delta_{0}}$.
We consider a Kolmogorov-type turbulence spectrum in proximity of the Galactic disk (with $\delta_{0}=1/3$) where the 
turbulence is expected to be injected by SNRs.
Away from the disk ($l<|z|<L$), the diffusion scaling index is taken as $\delta_{0}+\Delta$,
reflecting the effects of CR-driven turbulent motion. We use value $\Delta=0.55$ which is
inferred from primary CR spectra \citep{Feng2016}. 
The total production of secondaries in the ISM depends on the ratio $L/K_{0}$, which is taken as free parameter.
The near-Earth CR fluxes are affected by solar modulation.
We adopt the \textit{force-field} approximation using $\phi_{\rm Fisk}=0.7$\,GV as modulation potential for \AMS{} \citep{Ghelfi2016}. 
This value was derived using CR proton data.
Modulation of antinuclei can be different due to charge-sign dependent effects arising from drift motion.
We have investigated this difference \emph{a posteriori} using a 2D numerical model \citep{Kappl2016}.
However, we found no appreciable difference for the \AMS{} data, because these data
are collected across the 2013 polarity reversal, and thus they contain CR particles propagated under both polarity states.

\begin{figure}[!t]
  \centering 
  \includegraphics[width=0.975\textwidth]{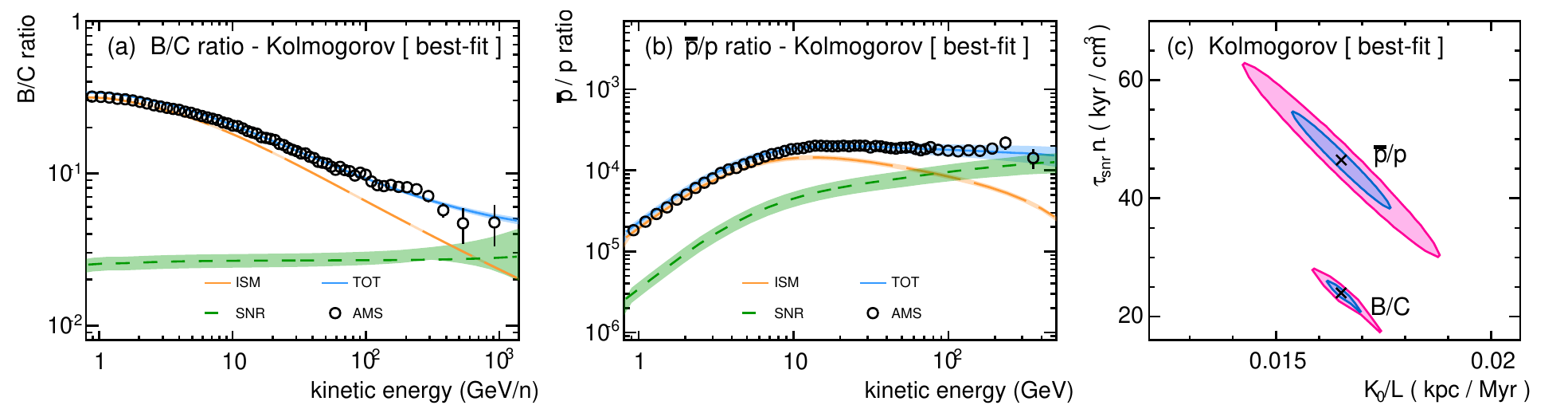}
\caption{\footnotesize%
  \BC{} ratio (a) and \pbarp{} ratio (b) measured by \AMS{} \citep{Aguilar2016BC,Aguilar2016PbarP}
  in comparison with best-fit models (light blue solid lines) with $E_{\rm min}=$\,10\,GeV/n.
  Contributing components from standard secondary production (orange, long-dashed lines) are shown together with the
  component from secondary production in SNRs (green short-dashed lines). The shaded bands represent the errors from the fits.
  Two-dimensional contour plots (c) are shown for the parameters $\tau_{\rm snr}n_{-}$ and $K_{0}/L$
  at 68\,\% (shaded blue area) and 95\,\% (shaded pink area) of confidence levels.
  The plots correspond to \BC-driven (a) and \pbarp-driven (b) fits.
}\label{Fig::ccSecPriRatiosFits}
\end{figure}

\section{Results: \BC{} and \pbarp{} ratios} 
\label{Sec::ResultsPbarBC}                   

We use the \AMS{} data on the \BC{} and \pbarp{} ratios to constrain key SNR parameters:
the product $\tau_{\rm snr}n_{-}$ between SNR age and upstream gas density, 
the ratio $K_{0}/L$ between normalization of diffusion coefficient and half-height of the halo,
and the normalization of the diffusion coefficient $D$ at the shock.
Fits are performed by means of a standard $\chi^{2}$ analysis, where \BC{} and \pbarp{} data are used separately.
The  $\chi^{2}$-estimator is calculated above the minimal energy threshold of 10\,GeV/n
in order to prevent possible biases from solar modulation. Other details will be provided in a forthcoming paper.
The best-fit models are shown in Fig.\,\ref{Fig::ccSecPriRatiosFits} for the \BC{} (a) and \pbarp{} (b) ratio
calculations along with the ISM and SNR contributions.
As seen from the figure, we have obtained very good fits for both observables, \ie,
accounting for secondary production inside SNRs improves the description of the data significantly.
The decrease of the ISM-induced \BC{} ratio is well balanced by the hard SNR-accelerated \B-component.
Furthermore, tor SNRs of typical age $\tau_{\rm snr}\cong\,20$\,kyr, the best-fit density is
close to the average ISM density $\tilde{n}\approx$\,1\,cm$^{-3}$. 
Unfortunately, the manifestation of the SNR component at high-energy does not provide striking signatures on the \BC{} ratio, but only a smooth hardening.
Furthermore, accounting for SNR-induced contributions in secondary CR fluxes introduces strong degeneracies between the source and transport parameters,
as discussed in \citep{TomassettiDonato2012,TomassettiDonato2015} using pre-\AMS{} data.
It was also noted that disregarding interactions in SNRs would favor models with weaker dependence for the Galactic diffusion coefficient.

Best-fit models for the \pbarp{} ratio are shown in Fig.\,\ref{Fig::ccSecPriRatiosFits}(b).
It can be noted that this ratio leads to a stronger secondary production in SNRs.
Nonetheless, \BC-driven and \pbarp-driven fits lead to different parameter values.
This can be seen in Fig.\,\ref{Fig::ccSecPriRatiosFits}(c), which shows the 68\,\% and the 95\,\% contours
for the parameters $K_{0}/L$ and $\tau_{\rm snr}n_{-}$.
The best-fit associated with the two observables lie in separate regions of the parameter space 
and, in particular, the \pbarp{} ratio favors denser media for the SNR background plasma.
This reflects the fact that, while the \BC{} data decreases steadily up to $E\sim\,$1\,TeV/n,
the \pbarp{} ratio is essentially constant at $E\gtrsim$\,50\,GeV.
This tension, that was noted under conventional models of CR propagation, is not 
resolved after accounting for secondary CR production and acceleration inside SNRs.

\section{Results: \dbar{} and \hebar{} fluxes}  
\label{Sec::ResultsAbar}                        

\begin{figure}[!t] 
\centering
\includegraphics[width=0.95\textwidth]{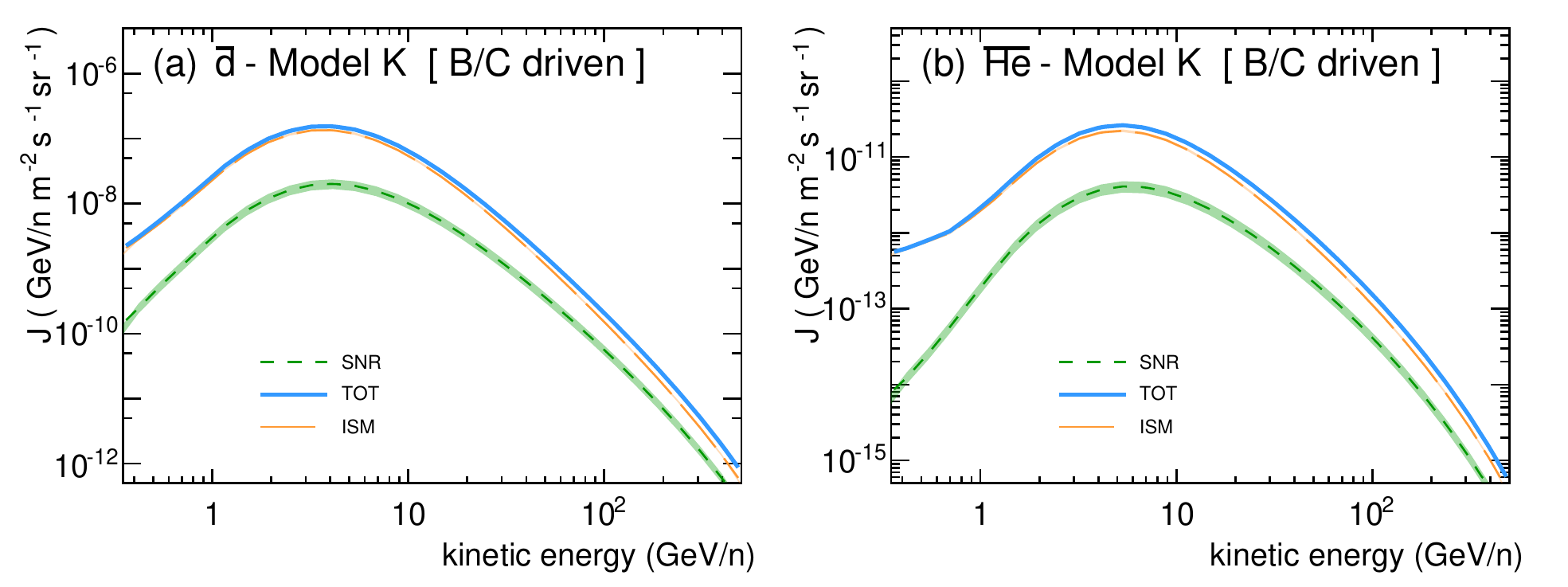}
\caption{\footnotesize%
  Model predictions of for the total flux of antideuterons (a) and antihelium (b), including secondary production inside SNRs 
  (green short-dashed lines), standard production in the ISM (orange long-dashed lines), and total flux (light-blue solid lines).
  Calculations are fromx the \BC-driven parameter setting.
}\label{Fig::ccAbarBCDriven}
\end{figure}

\begin{figure}[!t] 
\centering
\includegraphics[width=0.95\textwidth]{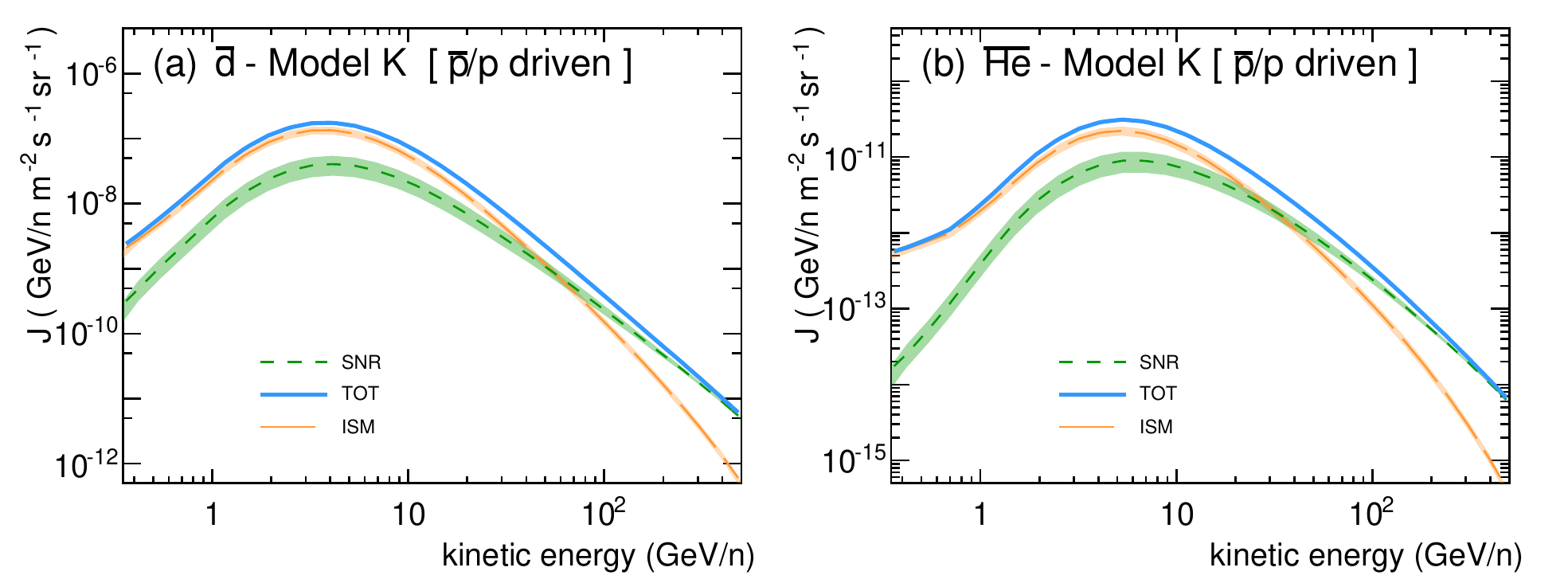}
\caption{\footnotesize%
  Model predictions of for the total flux of antideuterons (a) and antihelium (b), including secondary production inside SNRs 
  (green short-dashed lines), standard production in the ISM (orange long-dashed lines), and total flux (light-blue solid lines).
  Calculations are fromx the \BC-driven parameter setting.
}\label{Fig::ccAbarPbarPDriven}
\end{figure}

We now present our model predictions of antideuteron and antihelium energy spectra,
where  acceleration and propagation parameters are constrained by the \AMS{} data on \BC{} and \pbarp{} ratios.
The results are presented in Fig.\,\ref{Fig::ccAbarBCDriven} for the \dbar{} and \hebar{} fluxes near-Earth
from the \BC-driven parameter setting of Sect.\,\ref{Sec::ResultsPbarBC}.
The \dbar-fluxes are shown in the left panels and the \hebar-fluxes in the right ones.
Colors and line styles are encoded as in Fig.\,\ref{Fig::ccSecPriRatiosFits} so that,
along with the standard secondary ISM-induced component (orange),
the flux contribution from SNR-accelerated nuclei (green) is shown.
The shapes and the intensities of the fluxes are in good agreement with those reported in early works \citep{Donato2008}.
In particular, all particle fluxes show a characteristic peak at a few GeV/n of kinetic energy preceded and
followed by quick spectral drops at lower and higher energies. This shape reflects the kinematics of their
production and the rapid power-law falling flux of the progenitors, respectively. 
It can be  seen from the figure that the SNR flux contribution is always sub-dominant
over the considered energy range 0.5\,--\,500 GeV/n.
Under this model, however, SNR-accelerated secondaries do not appreciably contribute to the
total flux in low-energy region, although tertiary production processes have been included in the calculations.
In Fig.\,\ref{Fig::ccAbarPbarPDriven}, flux calculations of both species are presented for the \pbarp-driven parameter
settings. In contrast to the more conventional \BC-driven approach of Fig.\,\ref{Fig::ccAbarBCDriven},
\pbarp-driven predictions give an enhanced flux of SNR-accelerated antinuclei.
We also note that the \pbarp{} ratio is more tightly connected with the \dbar{} and \hebar{} spectra, because
all these antiparticles (\pbar, \dbar, \hebar) are generated from the fragmentation of the same CR species.

From calculations shown in Fig.\,\ref{Fig::ccAbarPbarPDriven}, it can be noticed that the SNR-accelerated
fluxes become dominating in the energy region above $\sim$\,100\,GeV/n.
At the energy of about $\sim$\,500\,GeV/n, the total fluxes of \dbar{} and \hebar{}
are one order of magnitude larger than those arising from \BC-driven calculations. 
In such a high-energy region the total flux is expected be very weak,
at the level of $10^{-11}$ ($10^{-14}$) GeV$^{-1}$\,m$^{-2}$\,$^{-1}$\,sr$^{-1}$ for \dbar{} (\hebar) particle.
This level is inaccessible by the existing CR detection experiments.
On the other hand, in the sub-GeV/n energy region, \pbarp-driven calculations
of Fig.\,\ref{Fig::ccAbarPbarPDriven} do not shown substantial differences in comparison with the \BC-driven
results of Fig.\,\ref{Fig::ccAbarBCDriven}.
At these energies, hadronic production processes in SNRs do not provoke significant modifications on the expected fluxes.
Furthermore, in spite of appreciable error bands on the SNR and ISM component,
the \emph{total} ISM+SNR flux prediction is found to be rather stable for a large region of parameter space.
\BC{} and \pbarp-driven prediction give both consistent results for the \emph{total} flux intensity of CR antinuclei
in the GeV/n energy region, event though the single SNR and ISM contributions appear to be substantially different.
In this respect, the total astrophysical background for DM searches appears to be soundly assessed at mid-low energies.
These consideration do not apply to the shape of DM induced signals.
The spectra of antinuclei produced from DM annihilations, in fact, suffer from different types of parameter degeneracy
because the DM source is expected to be distributed in the whole Galactic halo. Thus, when modeling
DM-induced CR particles, constaints provided by the \BC{} ratio are not sufficient to characterize their Galactic transport.
We also recall that the antinuclei flux calculations are affected by large nuclear uncertainties that we have not addressed in this work.
These uncertainties have a similar influence on ISM and SNR components, being the two contributions tightly correlated each other.

\section{Conclusions} 

This work is motivated by the search for a model of origin and propagation of Galactic CRs 
that is able to account for the several conflicting observations in their energy spectrum.
SNRs are thought to be the main sources of \emph{primary} CRs in the Galaxy such as proton, \He, or \C-\N-\Oxy{} nuclei.
In this work, we have used new data  from \AMS{} to determine the effect of production and acceleration
of \emph{secondary} CR particles inside these sources.
Consequence of this mechanism is a progressive flattening of secondary-to-primary ratios at high energies.
We have tested this mechanism under a two-halo scenario of CR propagation, based on Kolmogorov diffusion in the Galactic disk,
which provides an accurate description of the primary CR spectra. 
We found in the first place that the \AMS{} data support the production of secondaries in SNRs. 
Accounting for such processes, in fact, provides a good description of the \BC{} data in the GeV-TeV range.
Furthermore, the resulting amount of \B-nuclei produced by inside SNRs appears to be in line with naive expectations, as noted in \citep{Aloisio2015}.
Unfortunately, this mechanisms introduces new degeneracies between acceleration and transport parameters.
Our fits to the \pbarp{} ratio data from \AMS{} provides further evidence for secondary production in SNRs,
but
the \pbarp-driven parameters are found  to be inconsistent at 95\,\% CL with the \BC-driven paramters.
Hence, the tension between \BC{} and \pbarp{} ratios, already noted within conventional models of
CR propagation, cannot be resolved in terms of hadronic interaction inside CR accelerators.

With the constraints provided by the two ratios, we have presented 
calculations for the expected fluxes of \dbar{} and \hebar{} antinuclei in CRs.
As we have shown, the SNR flux component of CR antinuclei can be appreciably large at high energy.
However, in the low-energy region between $\sim$\,0.1 and a few GeV/n, which is where DM-induced signatures
have chances to exceed over the background, the SNR-accelerated flux is found to be always sub-dominant.
Furthermore, we found that the \emph{total} ISM+SNR flux of antinuclei
is rather stable for a large region of parameter configurations, event though the single
SNR and ISM contributions appear to be substantially different and highly model-dependent.
Thus the astrophysical background for DM searches in the low-energy region appear to be soundly assessed
in this energy region. 
The antideuteron and antihelium search is ongoing \AMS{} experiment, and soon, by the GAPS detection project \citep{GAPS}.
\\
\\
{\footnotesize%
We thank our colleagues of the AMS Collaboration for valuable discussions.
AO acknowledges CIEMAT, CDTI and SEIDI MINECO under grants ESP2015-71662-C2-(1-P) and MDM-2015-0509.
NT acknowledges support from MAtISSE.
This project has received funding from the European Union's Horizon 2020 research and innovation programme under the Marie Sklodowska-Curie grant agreement No 707543.
}

\end{document}